# Entanglement Telegraphed Communication Avoiding Light-speed Limitation by Hong-Ou-Mandel Effect

## by Nick Herbert
quanta@cruzio.com

ABSTRACT: Faster-than-light communication is possible via quantum entanglement in an EPR experiment if one can overcome nature's "quantum modesty" and gain knowledge concerning the quantum state of a single photon. Using a special EPR source that produces precisely timed *doublets* of EPR photons (biphotons), I show how one can gain additional information about the photon's polarization state--information not accessible using the conventional one-photon-at-a-time EPR setup. Using the Hong-Ou-Mandel effect that describes how two identical photons combine at a neutral beamsplitter (NBS), I show how Bob can obtain enough polarization information about his B photons to unambiguously ascertain the polarization basis of Alice's distant measurement device which she has deployed to measure her A photons. I am proposing, in short, that EPR + HOM = FTL. Version 2 concludes with a refutation of its FTL claim due to Lev Vaidman.

INTRODUCTION
In the canonical EPR setup a source of polarization correlated light sends photon pairs A and B in opposite direction to observers Alice and Bob who have a free choice to pose one of several complementary (mutually exclusive) questions concerning the state of the photons in their possessions. At first glance it looks like Alice and Bob can send signals faster than light, for, if say Alice decides to measure Plane polarization (two possibilities: Horizontal or Vertical) of her A photons then, as a result of quantum entanglement, Bob's distant B photons seem to (instantly?) acquire plane polarization states identical to the ones that Alice measures. Likewise if Alice chooses to measure Circular polarization (two possibilities: Right or Left) of her A photons, then Bob's distant B photons seem to (instantly?) acquire circularly polarized attributes. Thus it seems that Alice's decision--whether to measure PP (Plane polarization) or CP (Circular polarization) of her A photons can change the polarization state of Bob's distant B photons--and do so instantaneously, faster-than-light.

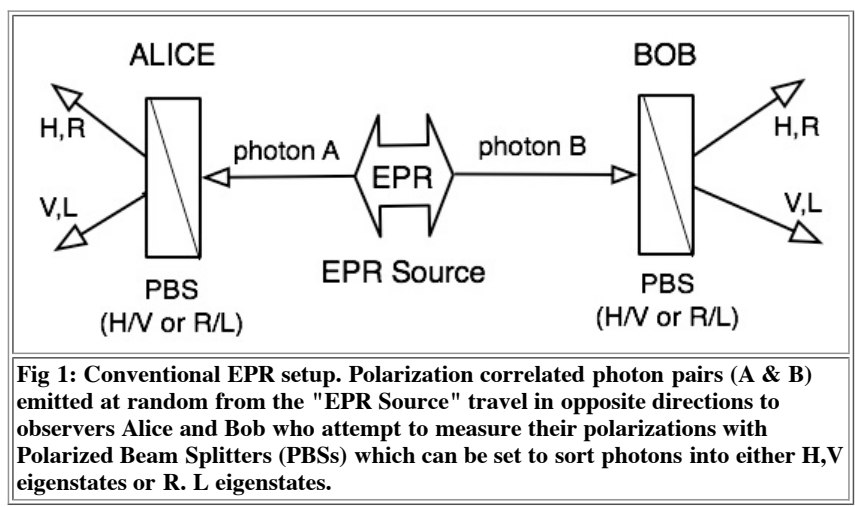

**Fig 1: Conventional EPR setup. Polarization correlated photon pairs (A & B) emitted at random from the "EPR Source" travel in opposite directions to observers Alice and Bob who attempt to measure their polarizations with Polarized Beam Splitters (PBSs) which can be set to sort photons into either H,V eigenstates or R. L eigenstates.**

However, things are not so simple. The EPR source (see Fig 1) produces paired photons in the state:

$$|EPR\rangle = |H(A)\rangle|H(B)\rangle - |V(A)\rangle|V(B)\rangle \ \ \ ................1)$$

These photons are 100% correlated in polarization but completely random. Thus if Alice decides to measure Plane polarization (by deploying a PBS--polarizing beam splitter--that splits light into H and V photons) she will observe a *random sequence* of Hs and Vs. In other words, her light looks completely unpolarized. On the other side of the world, Bob, if he should be deploying a similar H/V-style PBS, will obtain a random sequence of Hs and Vs that happen to be identical to the sequence of Hs and Vs at Alice's detectors. I call this sort of light--a random sequence of Hs and Vs--Plane Unpolarized light, or "PUP Light", for short.

Circularly polarized light consists of Right or Left CP photons where these states can be expressed thus as superpositions of H and V light.

$$|R\rangle = |H\rangle + i|V\rangle$$

$$|L\rangle = |H\rangle - i|V\rangle$$

Substituting these definitions into Eq 1) for the EPR state we obtain:

$$|EPR\rangle = |R(A)\rangle|R(B)\rangle - |L(A)\rangle|L(B)\rangle$$

which shows that, in an EPR state, CP light is also perfectly correlated. So if Alice chooses to measure CP light (by deploying a PBS-polarizing beam splitter--that splits light into R and L photons) then the polarization of Bob's photons, because of the perfect EPR correlations expressed above, will also change into R and L photons identical in sequence to the ones observed by Alice. Again these R and L photons, because of the nature of the EPR source, will be a random mixture. I call this sort of light--a random sequence of Rs and Ls--Circularly Unpolarized light or "CUP light", for short.

Now, one odd thing about quantum mechanics is that it describes these seemingly different kinds of light with the same density matrix. Quantum mechanics recognizes only one kind of polarized light. There is no such thing in QM as CUP light or PUP light. Both are simply called "unpolarized light". Period.

If Bob could distinguish CUP light from PUP light he could ascertain what Alice is doing at a distance, but quantum mechanics seems to forbid this distinction by describing both types of light as simply "unpolarized".

However there is one sense in which CUP light is certainly not "unpolarized". For Alice, when she produces Bob's CUP light, *knows* what the circular polarization of each of her photons is and can predict with 100% certainty what photon polarization Bob will observe should he choose to measure his B photons in the CP basis. Certainly if someone can predict the polarization of every single one of your photons, one would hesitate to call such a beam "unpolarized". But absent information from Alice, Bob's B photons appear to him to be randomly polarized no matter how he sets his Polarized Beam Splitters, either as H/V splitter or as R/L splitter.

However it would seem, because of Alice's complete predictive powers of Bob's photons for the particular choice of detector that Alice makes, that these CUP photons (if Alice chooses a R/L beamsplitter) do indeed possess some hidden property that Bob might determine if he were sufficiently ingenious. If by any means, Bob could find a way to tell the difference betwen CUP light and PUP light, then Alice could send Bob signals faster than light.

If CUP light and PUP light beams truly exist as "elements of reality" and yet fail to be described in the quantum formalism either as pure states or as mixtures, then perhaps Einstein was correct when he suggested that quantum mechanics is incomplete in the sense that there are real features of nature that escape its descriptive net.

## QUANTUM MODESTY

A quantum system is said to be in an "eigenstate" of a certain observable if when you measure that observable you get with 100% certainty a particular value (called the "eigenvalue" of that observable). For example if I prepare a photon in the H polarization state it is said to be in an eigenstate of the Plane polarization observable with eigenvalue "H".

In this context Einstein spoke of "elements of reality". He defined an "element of reality" thus: "If, without disturbing a system, you can predict with 100% certainty what the outcome of a measurement of some attribute of that system will be, then that attribute is an "element of reality.""

I make an H-polarized photon and hand it to you. In Einstein's sense my H-polarized photon possesses an "element of reality". I can predict with 100% certainty the outcome of your Plane-polarization measurement; I know what the polarization of this photon is because I created it that way.

Likewise when Alice measures Circularly polarized photons she "creates" ("discovers?") "elements of reality" at Bob's measuring site because (without disturbing his B photons) she can predict with 100% certainty the results of every CP measurement Bob might make on his photons.

So certain photons seem to possess "secret attributes" known to some but not to others. And if these attributes can be known (by others than their creators) the way seems open (via the EPR experiment) to transmit signals faster than light.

What are these hidden attributes? They are the eigenvalues of certain observables. They are Einstein's "elements of reality".

And there is general agreement among physicists that they are forever unknowable. By other than their creators.

For example, I know with certainty that the photon I am handing you is H polarized because I created it that way. But there is no way in Heaven and on Earth that you can discover this truth. Unless I tell you.

It's extremely odd that such a powerful principle believed (by most physicists) to be universally true--the unknowability of the eigenvalue of a single quantum--does not seem to have ever been given a name. Here I remedy that lack.

That a quantum system can possess a certain attribute with 100% certainty and yet this attribute is also 100% unknowable I will call "quantum modesty". Such a quantum system--all quantum systems--shyly hide "what she's really like" under an (allegedly) fully opaque veil of unmeasurability.

If you could find a way to measure the eigenvalue of the H photon I just handed you, or the eigenvalues of Alice's CUP light that she has distantly prepared for Bob, why then all Hell would break loose. Faster Than Light signalling would be possible. And also time machines.

So measuring a single system's eigenvalue had better be forever impossible. Or ordinary causality is in real trouble.

However I can discover nothing in quantum theory that would prohibit the discovery of a single system's eigenstate. If I can know (for sure) the polarization of a photon I just handed you, why cannot you somehow discover it? It can't be Totally Forbidden Knowledge because I (for one, and Alice, for another) possess it.

There are certainly many things that quantum mechanics with good reason says that it is impossible to know. For instance the momentum and position of a quantum system cannot be known, not because of any limitation on human measuring ability, but because such attributes DO NOT EXIST. There is no quantum system that possesses simultaneously both a momentum and a position eigenvalue, so it is nonsense to speak of making such a measurement. Like attempting to measure the area of a square circle. Likewise no photon can be in a simultaneous eigenstate of Plane and Circular polarization. These polarization bases are "conjugate attributes" like position and momentum. So when I hand you an H polarized photon (whose Plane polarization I know for certain) I can tell you nothing whatsoever about this photon's Circularly polarized attributes. An H-polarized photon has no "elements of reality" that correspond to the observable "Circular polarization". Likewise a Circularly-polarized photon has no elements of reality corresponding to the observable "Plane polarization".

So at first glance I cannot find within quantum mechanics any justification for attributing permanent status to nature's "quantum modesty". The information is certainly there. Why can't I know it? Why can't I slip inside her veils and sense, at least in part, the hidden beauty that I know for sure certainly exists underneath? Why so coy, O beautiful quantum world?

The rest of this article is concerned with a novel proposal to peek behind the veils of "quantum modesty". It's an attempt to modify the simple EPR experiment of Fig 1 so that Bob can actually measure the kind of eigenstates that Alice is distantly preparing for him. It's a proposal that--if it works--will allow Bob to detect a difference between CUP and PUP light--a difference that quantum mechanics says is not there. Or if it is, it's protected by nature's quantum modesty. It's a difference that Alice can certainly demonstrate to Bob but only via conventional lightspeed-limited communication. Let's see how Bob might attempt to discover his photon's hidden "random" polarization sequence without help from Alice.

THE ENTANGLED BIPHOTON MACHINE--A MODIFIED EPR EXPERIMENT

To achieve an extra degree of experimental control we first replace the usual EPR light source which emits entangled photon pairs at random times with a source that produces these pairs in a particular pattern illustrated in Fig 2. Two photon pairs m and n are emitted separated by a constant time T(d). These photon doublets are separated by a time T(r). In practice the rhythm of the source is two beats, pause; two beats, pause. The photons that travel to Alice are left alone. The photons m and n that travel to Bob are manipulated with mirrors so that both photons in the doublet arrive at Bob's neutral beam splitter (NBS) simultaneously (within the same coherence time). See Fig. 3 for details of how photons m and n are temporally superimposed using a switchable mirror to delay "early photon" m so that its arrival time at Bob's matches that of "late" photon n.

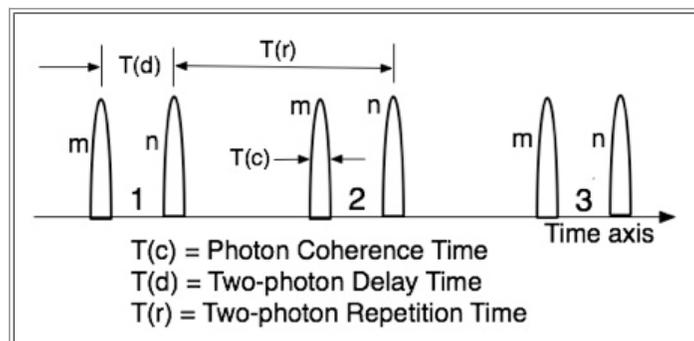

Fig. 2: The temporal structure of the EPR source for the Entangled Biphoton Machine. Two pulses m and n are produced separated by a delay time T(d), then a pause. Then two more pulses separated by the same time interval T(d). The time between pulse doublets is T(r), both intervals long compared to the photon coherence time T(c).

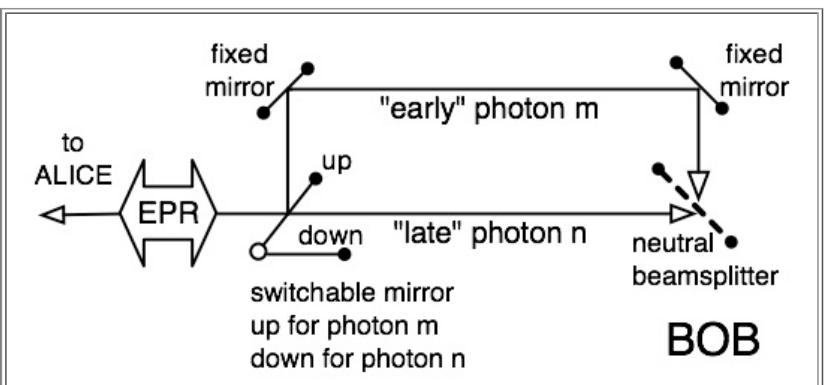

Fig. 3: A switchable mirror reflects the "early" photon m into a longer path than the "late" photon n. The longer path is adjusted so that both photons reach Bob's neutral beam splitter (polarization insensitive 50/50 mirror) at the same time.

Now the fun begins. Bob sets out to measure the polarization state of his Alice-modified photons. But instead of measuring one photon at a time he measures two simultaneous photons that have been united (or separated) by a polarization-insensitive beamsplitter with 50/50 transmission--the so-called "half-silvered mirror".

BOB MEASURES BIPHOTONS CREATED FROM PUP LIGHT

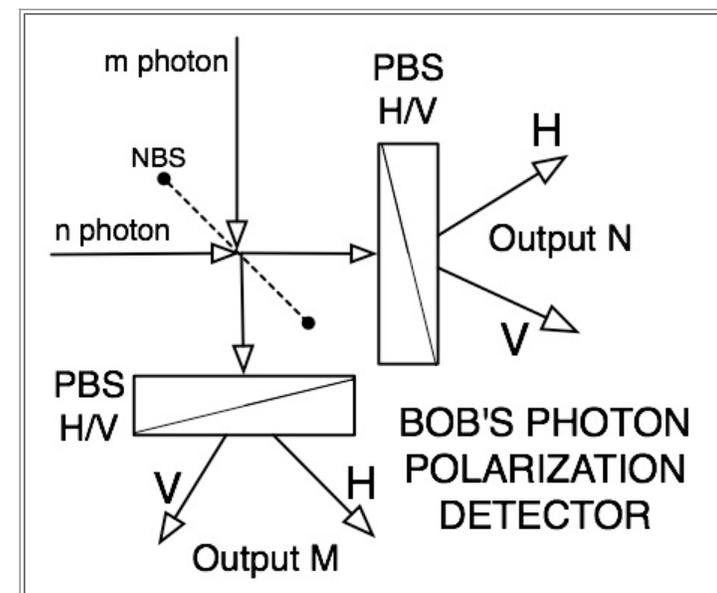

When two identical photons impinge on a neutral beamsplitter (NBS) they are subject to the Hong-Ou-Mandel Effect which requires them to both enter the SAME channel of the NBS (1). In Bob's case, if photons m and n are identical, they must both go into NBS channel M or NBS channel N. The gist of the HOM Effect is that, at a beamsplitter, identical photons don't split. Identical photon always stick together.

On the other hand, if photons m and n are not identical they are free to split or to stick together and do so at random--half the time they stick together and enter the same NBS channel and half the time the two photons split and enter two different channels. The statistical result of two different photons impinging on a NBS resembles the tossing of two pennies where heads and tails represent the two output channels M and N. For the two pennies, there exist four equally likely outcomes: HH, TT, HT and TH; likewise for the two non-identical photons: channels MM, NN, MN and NM are the four equally likely outcomes.

The behavior of identical photons impinging on a neutral beamsplitter summarized above is the key to understanding the behavior of this faster than light communication device. Please be sure you clearly understand the content of the last two paragraphs before proceeding further. These last two paragraphs contain the key to the human conquest of time, space and matter. Time spent rereading these few important sentences will not be wasted.

Now that we are thoroughly familiar with the behavior of two photons entering a beamsplitter, I will show how Bob can lift nature's veil a bit and determine the "secret elements of reality" that dear distant Alice has prepared for him. Armed with the above knowledge I will show how Bob can easily distinguish

Fig. 4: The m and n photons are simultaneously united at NBS, the neutral beam splitter. For each photon pair there exist only two possibilities--either both photons go into the SAME channel (both into channel M, for instance) or each photon goes into a DIFFERENT channel--one into M, the other into N. It is these latter photons that will be used for measurement. Photons that go into the same channel are ignored.

PUP light from CUP light, hence "crack the cosmic code" that (until this day) has prevented humans from signalling FTL.

At each output (M and N) of the NBS, Bob places a Polarizing Beam Splitter (PBS) which sorts photons into two orthogonal polarization channels. Bob has a choice of polarization bases--either he can divide the beam into H or V photons (PBS H/V) or he can divide the beam into R and L photons (PBS R/L). He chooses (see Fig. 4) to place H/V beamsplitters in both channel M and channel N and to record their outputs.

Suppose Alice is using a PBS H/V to measure her distant A photons. This means that she is observing a particular sequence of H and V photons. And because of EPR entanglement, the same sequence of H and V photons is appearing in Bob's vicinity. By her actions Alice is causing PUP light (rather than CUP light) to appear at Bob's detector.

PUP light consists of a random mixture of Plane-polarized photons--a random sequence of Hs and Vs. But our new improved EPR light source produces for Bob a series of "biphotons" by merging together two time-adjacent photons. So Bob is not measuring a random sequence of single photons (H, V, V, H, V, H, H, V... for instance) but a random sequence of bipolarizations (HH, HV, VH, VH, VV, HH...for instance). Because of this new type of EPR light source, Bob has four letters in his crypto message to decode rather than a two-character code as in conventional EPR experiments. Let's examine the fate of each of these four plane-polarization code letters as it traverses Bob's detector.

**HH**: When an H photon hits both inputs of the NBS, the HOM Effect forces both photons to enter the same output channel, either channel M or channel N, the outcomes labeled MM and NN in Table 1. The important feature here is that no H photons enter both channels, indicated by the "0" entries in Table 1.

**VV**: Similar to the HH biphoton, the HOM Effect confines these V photons to the same NBS channel.

**HV**: Since these photons are not identical, HOM does not apply. The result resembles flipping two coins: half the time the two photons end up in the same channel; and half the time they split, ending up in two different channels.

**VH**: Same behavior as HV: half stick together; half split. The behavior of each of these four biphotons is summarized in Table 1.

|  | MM | NN | MN | NM |
|---|---|---|---|---|
| H(m)H(n) | HH | HH | 0 | 0 |
| V(m)V(n) | VV | VV | 0 | 0 |
| H(m)V(n) | HV | VH | HV | VH |
| V(m)H(n) | HV | VH | HV | VH |

Table 1: Outcome of Bob's detectors for PUP light input.

Table 1 not only describes the behavior of the biphotons on the neutral beamsplitter. But also, because Bob has deployed Plane-polarized beam splitters in both channels (M and N), Table 1 describes the outcome of Bob's four photon detectors. Bob now proceeds to read Alice's coded message. He proceeds in two steps.

**Step 1**: At each polarization detector two possibilities exist: either the detector is triggered by two photons or by only one. Bob's first step is to IGNORE ALL CASES where two photons enter the same detector (call this the "2-0 response"). This means that all responses on the left half of Table 1 are ignored. Only the right half of Table 1 is of interest for FTL communication. In this step Bob throws away 75% of his data; only 25% of the photon pairs are used for FTL signalling.

**Step 2**: When Bob examines the right half of Table 1, he notices that in the cases where one photon enters NBS channel M and the other enters NBS channel N (call this the "1-1 response"), that THESE PHOTONS ARE ALWAYS DIFFERENT. When biphotons made from PUP light split they always possess opposite polarizations.

|  | H(M) | V(M) |
|---|---|---|
| H(N) | 0 | 50% |
| V(N) | 50% | 0 |

Table 2: Response of 1-1 PP detectors to PUP Light

Focusing only on the cases (1-1) where the PUP light photons enter different polarimeters we summarize the photon counting results in Table 2. We see that never when a photon "splits" do the two photons go into the same Plane polarization channel.

I claim that this behavior constitutes an unambiguous signature for the presence of PUP light--consequently a de facto demonstration of superluminal communication--a readable message sent FTL from Alice to Bob.

BOB MEASURES BIPHOTONS CREATED FROM CUP LIGHT

Because of the results of his measurements (summarized in Table 1) Bob believes that he now has evidence that Alice has been sending him PUP light. Suppose Alice changes her mind. And decides to send CUP light instead. Can Bob determine (faster than light) that far-away Alice has changed her mind?

Let's see.

The HOM Effect is not concerned with the actual attributes of photons--only whether these attributes are the same or different. So the response matrix for CUP biphotons looks exactly the same (Table 3) as the response matrix for PUP biphotons with the simple substitution H = R, V = L. The essence of the HOM effect is that similar photons are clumped, dissimilar photons are randomly distributed (clumped and unclumped).

As before, Bob entirely ignores the left half of the response matrix--which describes only photons that take the same path (which we called 2-0 type)--and directs his attention to the right half of the response matrix which reports that paired photons of the 1-1 variety are always DIFFERENT, either RL or LR.

|  | MM | NN | MN | NM |
|---|---|---|---|---|
| R(m)R(n) | RR | RR | 0 | 0 |
| L(m)L(n) | LL | LL | 0 | 0 |
| R(m)L(n) | RL | LR | RL | LR |
| L(m)R(n) | RL | LR | RL | LR |

Bob could have verified directly the presence of the RL photons if he had deployed R/L Polarized beam splitters in channels M and N. But he had chosen instead to deploy H/V beamsplitters. Because an R photon is in a sense "half H, half V" it will respond probabilistically to a H/V beamsplitter and trigger 50/50 randomly the H and V channels of this detector. So a collection of RL and LR photons that split and enter the two H/V polarized beam splitters will produce the experimental signature of Table 4.

**Table 3: Output of Bob's detectors for CUP light input.**

|      | H(M) | V(M) |
|------|------|------|
| H(N) | 25%  | 25%  |
| V(N) | 25%  | 25%  |

**Table 4: Response of 1-1 PP detectors to CUP light**

The biphotons that split, in this case, and go into different channels are RL biphotons whose nature is to register 50/50 randomly as either an H or a V photon. Thus when Alice has deployed her H/V detector (which produces PUP light for Bob to look at), Bob observes the pattern of 1-1 PP detections depicted in Table 2. But when Alice deploys her R/L detectors (which produces CUP light for Bob to look at), Bob observes the pattern of 1-1 detection events displayed in Table 4. Since the statistical patterns of Table 2 and of Table 4 are decidedly different one can only conclude that Bob is able to experimentally determine the difference between PUP and CUP light, an ability which permits Alice to send Bob a message faster than light. Is there a time machine in our future? And who will bring it to market first, Sony or Apple Computer?

## CONCLUSION

This scheme for sending superluminal signals from Alice to Bob depends on the ability to look not at single polarized photons (H, V, R, L) but to construct a more expanded polarization alphabet made of biphotons (HH, HV, RL, etc) instead. The present proposal is based on an EPR source with a precisely timed pulse structure--a source which is certainly impractical to build with present technology but is not forbidden by the laws of physics. This scheme represents then a thought experiment--in the spirit of the original EPR experiment (1935) which took almost 50 years to realize in practice. The double-pulse EPR source envisioned in this paper was chosen for simplicity of explication. Actually any scheme which can determine the emission time of the EPR photon pairs can be utilized in principle to construct the biphotons necessary for the operation of this superluminal signalling scheme. This scheme succeeds because it incorporates one additional piece of information lacking in all previous attempts to exploit EPR for FTL. That additional piece of information is the emission times of the correlated EPR pairs.

For instance, given the time of emission of each EPR photon, one can employ a precisely timed source of H-polarized photons to mix a H photon with the unknown EPR photon at the NBS. One has then the task of distinguishing the two-letter alphabet HH, HV in the PUP light case from the two-letter alphabet HR, HL in the CUP light case. The total absense of HH coincidences in the former case would again provide an unmistakable signature for the presence of PUP light. From an engineering standpoint it would certainly be easier to build a source of H photons whose emission time can be precisely controlled, then to build an EPR source whose emissions can be controllably pulsed.

Or, once in possession of the emission time of each EPR photon, one could construct a custom delay interval for each consecutive photon doublet so that each member of the doublet arrives at Bob's mirror at the same time--a sort of asynchronous variation of the present proposal.

Or, we could renounce precise timing entirely and exploit chance coincidence to get our EPR photons to meet at the same mirror. Instead of switching single photons, the switchable mirror of Fig 3 would deflect an entire photon sequence of length T and would delay this sequence exactly T seconds so that the front of this (deflected) m beam reaches the mirror precisely coincident with the beginning of the undeflected n beam. Then chance coincidence will insure that a certain fraction of the photons enter the mirror simultaneously. The rate of coincidence K of two beams of photon rate N and coincidence time $T(x)$ is given by $K = N \times N \times T(x)$. So for EPR photon rates of N = 1 megacycle and a coincidence time of 1 nanosecond, the rate K of accidental coincidences is 1 kilocycle. Assuming we will need about 100 coincidences for a reliable measurement, this suggests a realistic data rate of about 10 bits per second. This bit rate may seem small, but this scheme uses only simple well-established technology, strongly suggesting that the task of coaxing independent EPR photons to meet at the same mirror is not the crucial bottleneck in this new FTL proposal.

None of these schemes for persuading two consecutive EPR photons to meet at the same mirror seems to violate any fundamental physics principles.

Once a mechanism is in place for producing biphotons for Bob, the rest of the detection scheme depends only on off-the-shelf available hardware--neutral beam splitters, polarized beam splitters and single-photon detectors. No cumbersome, messy gain tubes, for instance, as in my previous FLASH FTL proposal (2).

Cutting through the detail, this superluminal signaling scheme is remarkably simple. The task is to distinguish which of two random alphabets is present in Bob's beam--either a four-letter PUP alphabet consisting of the characters HH, VV, HV and VH, or a CUP alphabet consisting of the letters RR, LL, RL and LR. The HOM effect and the restricted attention only to 1-1 events "miraculously" filters out all but two kinds of events--#1: events in which an HV biphoton splits and goes to two different polarization-sensitive detectors and #2: events in which an RL biphoton splits and goes to two different polarization-sensitive detectors. But these two cases are quite easy to distinguish experimentally. Hence Bob can experimentally make the PUP/CUP decision and receive Alice's messages instantly. Now that's True Instant Messaging!

One minor drawback of this scheme is that it is somewhat inefficient--only 25% of the biphotons are used for signaling. The rest are tossed out. This means that, in order to reliably send a single bit, Alice has to send a large number of similar transmissions in order to insure that Bob receives her message. But the signal to noise ratio is not too bad. If Alice designs her messages so that one bit is contained in a block of 64 pulse pairs, then Bob will receive 16 pairs which contain a useful signal. In this case, the odds are small that Bob will fail to detect the correct bit that Alice intended to send.

I have demonstrated here a workable superluminal signaling scheme based on constructing and detecting polarized biphotons from an EPR source. All that remains is to give it a name. For that purpose the awkward title of this paper forms an acronym: Entanglement Telegraphed Communication Avoiding Light-speed Limitation by Hong Ou Mandel Effect.

Henceforth this scheme, whether successful or not, shall be known as: ETCALLHOME.

## REVISED VERSION: REFUTATION OF THE PROPOSED BIPHOTON FTL SIGNALING SCHEME

One of the crucial assumptions of this proposed FTL communication scheme concerns the behavior of R and L photons at a neutral beamsplitter when CUP light is being detected. I believed, that since these are orthogonal states, that they would split independently of one another and that an equal mixture of R(M)L(N) and L(M)R(N) states would appear as an input to the H/V Polarized Beam Splitters. Since both R and L states are half H and half V, one should expect equal triggering of the four possibilities H(M)H(N), V(M)V(N), H(M)V(N), V(M)H(N)--like flipping a coin. This conjectured uniform pattern of outcomes is illustrated in Fig 4. Since Fig 4 (the conjectured CUP outcome) is different from Fig 2 (the conjectured PUP outcome), I concluded that this scheme is able to distinguish CUP from PUP light.

However, after this paper had been submitted to the arXiv, a spirited email exchange with Lev Vaidman convinced me that my picture of the behavior of R and L photons at a neutral beamsplitter was mistaken. Instead of the random mixture of RL, LR photons that I envisioned, the correct (1-1) output of the NBS is not a random mixture but an *entangled pair* of RL photons:

$$R(m)L(n) \longrightarrow R(M)L(N) - L(M)R(N) \dots\dots\dots\dots\dots\dots\dots\dots\dots 2)$$

When this entangled pair is decomposed into H and V photons, a remarkable cancellation occurs that removes the diagonal terms HH and VV:

$$(H(M) + iV(M))(H(N)-iV(N)) - (H(M) - iV(M))(H(N) + iV(N)) = 2(V(M)H(N) - H(M)V(N)) \dots\dots 3)$$

Thus at Bob's detectors only the off-diagonal events H(M)V(N) and H(N)V(M) are triggered and in equal measure. (No HH or VV events occur.) Consequently Fig 4 becomes the same as Fig 2. Conclusion? This scheme cannot distinguish CUP and PUP light from one another. This FTL signaling scheme fails.

In retrospect, it seems ironic that a scheme to achieve FTL signalling via the existence of non-local photon entanglement between Alice and Bob (Eq 1) should be derailed by the existence of local photon entanglement (Eq2) at Bob's detector.

The message? That this is probably not the way to woo her. Even granted access to EPR photon emission times, Bob is unable to lift nature's veils. Her quantum modesty remains intact.

NICK HERBERT
quanta@cruzio.com
December 6, 2007
Feast of Saint Nicholas

I am pleased to dedicate this paper to the memory of Heinz Pagels who believed it to be impossible.
I would like to thank Lev Vaidman for his gentle tutelage in the art of photon algebra.

## REFERENCES
(1) C. K. Hong, Z. Y. Ou, L. Mandel Phys Rev Letters **59** 2044 (1987)
(2) N. Herbert Foundations of Physics **12** 1171 (1982)

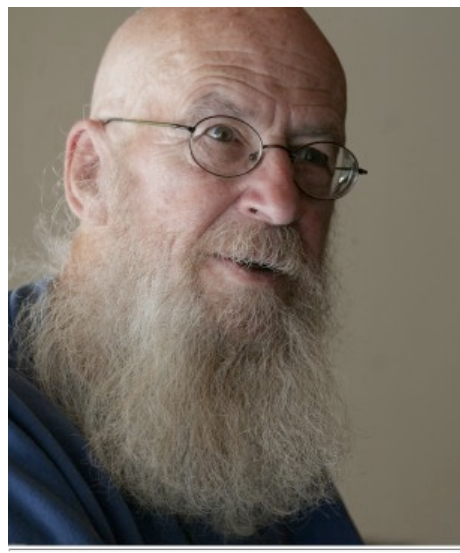

NICK HERBERT is the author of "Quantum Reality",
"Faster Than Light", "Elemental Mind" and a chapbook
"Physics on All Fours". He devised the shortest proof
of Bell's Theorem, had a hand in the Quantum No-Cloning
Rule and is presently obsessed with Quantum Tantra. He
will now henceforth be associated with ETCALLHOME.